\documentstyle[prl,aps,preprint,tighten,floats]{revtex}

\begin{document}
\def \sin {{\rm sin}}
\def \cos {\ {\rm cos}}
\def \tan {{\rm tan}}
\def \cot {{\rm cot}}

\def \non{\nonumber}
\def \wzw {$SL(2,R)\times SU(2)\ $}
\def \ads {$AdS_3 \times S^3\ $}
\def \de {\del}
\def \A {{\cal A}}
\def \AA {{ \cal B}}

\newcommand{\rf}[1]{(\ref{#1})}

\def \foot {\footnote}
\def \bi{\bibitem}
\def \la {\label}
\def \tr {{\rm tr}}
\def \ha {{1 \over 2}}

\def\np {{Nucl. Phys. }}
\def \pl {{Phys. Lett. }}
\def \mpl {{Mod. Phys. Lett. }}
\def \prl {{Phys. Rev. Lett. }}
\def \pr  {{Phys. Rev. }}
\def \ap  {{\em Ann. Phys. }}
\def \cmp {{\em Commun.Math.Phys. }}
\def \ijmp {{\em Int. J. Mod. Phys. }}
\def \jmp {{\em J. Math. Phys.}}
\def \cqg {{Class. Quant. Grav. }}

\def \del{\partial}
\def \m {\mu}
\def \n {\def \nu}
\def \ci {\cite}
\def \g {\gamma}
\def \G {\Gamma}
\def \ka {\kappa}

\def \A {{\cal A}}
\def \AA {\td {\cal A}}
\def \third {{\textstyle {1\ov 3} } }
\def\Jo#1#2#3#4{{#1} {\bf #2}, #3 (#4)}
\def \re#1{(\ref{#1})}
\def\st{\scriptstyle}
\def\sst{\scriptscriptstyle}
\def\mco{\multicolumn}
\def\epp{\epsilon^{\prime}}
\def\vep{\varepsilon}
\def\ra{\rightarrow}
\def\vp{{\bf p}}
\def\al{\alpha}
\def\ab{\bar{\alpha}}
\def \bi{\bibitem}
\def \ep{\epsilon}
\def\D{\Delta}
\def\sms{$\s$-models }
\def \om {\omega}
\def \foot{\footnote}
\def\be{\begin{equation}}
\def\ee{\end{equation}}
\def \lab {\label}
\def \k {\kappa} 
\def \F {{\cal F}}
\def \g {\gamma}
\def \del {\partial}
\def \bd {\bar \partial }
\def \na {\nabla}
\def \const {{\rm const}}
\def \ha{{\textstyle{1\over 2}}}
\def \na {\nabla }
\def \D {\Delta}
\def \a {\alpha}
\def \b {\beta}
\def \chi {\chi}\def\r {\rho}
\def \s {\sigma}
\def \p {\phi}
\def \m {\mu}
\def \n {\nu}
\def \vp {\varphi }
\def \l {\lambda}
\def \t {\theta}
\def \td {\tilde }
\def \d {\delta}
\def \ci {\cite}
\def \la {\label}
\def \sm {sigma-model }
\def \foot {\footnote }
\def \P {\Phi}
\def \o {\omega}
\def \inv {^{-1}}
\def \ov {\over }
\def \four{{\textstyle{1\over 4}}}
\def \fourth{{{1\over 4}}}
\def \foot{\footnote}
\def\be{\begin{equation}}
\def\ee{\end{equation}}
\def\bea{\begin{eqnarray}}
\def\eea{\end{eqnarray}}
\def\np {{\em  Nucl. Phys. }}
\def \pl {{\em  Phys. Lett. }}
\def \mpl {{\em Mod. Phys. Lett. }}
\def \prl {{ \em  Phys. Rev. Lett. }}
\def \pr  {{\em  Phys. Rev. }}
\def \ap  {{\em Ann. Phys. }}
\def \cmp {{\em Commun.Math.Phys. }}
\def \ijmp {{\em Int. J. Mod. Phys. }}
\def \jmp {{\em J. Math. Phys.}}
\def \cqg {{\em  Class. Quant. Grav. }}

\date{\today}
\preprint{\vbox{\baselineskip=12pt
\rightline{Imperial/TP/97-98/55 }
\vskip0.2truecm
\rightline{UPR-0809-T}
\vskip0.2truecm
\rightline{ITP/NSF-98-069}
\vskip0.2truecm
\rightline{hep-th/9806141}}}
\title{
Sigma Model of 
Near-Extreme Rotating\\ Black Holes 
 and Their
Microstates} 
\author{ Mirjam Cveti\v c${}^{\star\ \sharp}$\  and \ Arkady A.
Tseytlin${}^{\dagger}$~\footnote{Also at  Lebedev Institute, Moscow.}
}
\address{${}^\star$Department of Physics and Astronomy \\ 
University of Pennsylvania, Philadelphia PA 19104-6396, USA\\
${}^{\sharp}$Institute for Theoretical Physics\\
University of California, Santa Barbara, CA 93106, USA\\
${}^{\dagger}$Theoretical Physics Group, Blackett Laboratory\\
Imperial College,   London SW7 2BZ, U.K.}
\maketitle
\begin{abstract}
Five-dimensional non-extreme rotating black holes with large
NS-NS five-brane and fundamental string charge are shown 
to be described by a conformal sigma model, which is a 
marginal integrable deformation of six-dimensional 
$SL(2,R) \times SU(2)$  WZW  model. The two WZW levels
are equal to the five-brane charge, while the 
parameters of the two marginal deformations generated by the  
left and right chiral $SU(2)$ currents are proportional to the 
two angular momentum components of the black hole. The near-horizon description is  effectively in terms of a free fundamental string  whose tension is rescaled by the five-brane charge. 
The microstates are identified  with those of left and right
moving superconformal string oscillations in the four directions 
transverse to the five-brane. Their statistical entropy reproduces
precisely the Bekenstein-Hawking  entropy of the rotating black hole.  
\end{abstract} 
\pacs{\tt PACS number(s): }
\maketitle

\section{Introduction}
String theory, as a quantum theory of gravity,  should provide a framework 
within  which  one should be able to 
address quantum aspects of strong gravitational fields, in particular the issue
of  black hole information loss and related 
issues of  microscopic structure of black holes.  
 While the  information
loss is still an open unresolved problem, an important progress has been made in
clarifying   the black hole microscopics.   In the days of perturbative string
physics Sen made a pioneering 
proposal~\cite{sen95} to identify the microstates of extreme (BPS-saturated) 
electrically charged black
holes  with perturbative excitations of string theory. 
 This idea was clarified and  put on a firmer ground 
 in \ci{cmp,harv} by 
interpreting   the  extreme electric 
 black hole states as oscillating modes of 
underlying macroscopic string.

However,  only after the discovery of
    BPS-saturated
multi-charge dyonic black holes with regular
  horizons~\cite{kall,dyon}  and thus   finite
Bekenstein-Hawking (BH) entropy,  that attempts  to  establish  a  
quantitative agreement between the microscopic and macroscopic 
entropy  in string theory became feasible.  Such black holes  are necessarily
non-perturbative objects, and were originally  specified
as solutions of effective  four-dimensional   toroidally compactified 
string theory  with  charges from the  Neveu-Schwarz--Neveu-Schwarz (NS-NS)
sector~\cite{dyon,us1}. Their microscopic features 
were captured by string 
theory  in  {\it curved space-time geometry  of the 
near-horizon region}  which is that of 
an  $SL(2,{R}) \times SU(2)$
Wess-Zumino-Witten (WZW) model~\cite{lowe94,us2}.
In particular,  the small-scale  string oscillations~\cite{LW,us2}  were shown 
to reproduce
~\cite{us2,me1,me2} the extreme black 
hole entropy directly from the near-horizon geometry.
The key point  was  that  the BH entropy should represent 
the leading term in the  statistical entropy in the  limit  of 
large  charges 
(in which the characteristic scale of 
black hole is large and thus string corrections can be ignored). For large 
charges it  should be  sufficient to 
consider only the  
near-horizon 
region where the corresponding string sigma-model  takes the  WZW form.

It  was suggested in \ci{LW}  that 
 the BH entropy  of these
 BPS-saturated black holes can be  interpreted as the  
  entropy   of  {\it left-moving} supersymmetric
oscillating  states of a free string  with  tension 
renormalised by the (product of)  ``magnetic''  charges, i.e. 
non-perturbative charges of the NS-NS sector (solitonic five-brane and
Kaluza-Klein monopole charges).
The origin of  this  renormalisation 
was  explained in \ci{us1,me1}   
  starting with  the  conformal sigma-model 
which describes  the embedding~\ci{us1,us2,me1}   of the dyonic
black holes into string theory.  The  marginal supersymmetric deformations in
the left-moving sector 
of the  conformal \sm  were interpreted, in the spirit  of \ci{cmp,harv},   as 
microstates describing   degenerate black holes 
with the same asymptotic charges but different short distance 
structure. Since these perturbations  are important  only
at small scales  they can be effectively counted  near 
the horizon ($r\sim r_+$). The crucial  observation~\cite{us2}
 is that for  $r\to r_+$ the  model
 reduces to  a  six-dimensional WZW  theory with level
proportional to (product of) magnetic charges, so that for large charges
 its spectrum is thus
effectively that of the 
free fundamental string with  tension rescaled by the 
magnetic charges.
 While the subleading terms 
in the statistical entropy may  depend 
on embedding of the black hole into a particular string theory, 
the leading   term  should  be universal.

As was explained in \ci{me2},  the leading term receives contributions only 
from  certain 
 universal types of  perturbations which are common to 
heterotic and type II embeddings. Other perturbations, in particular,  
 string  oscillations in internal  toroidal   directions, which do not carry
 charges,
 contribute  only to {subleading} terms in the entropy.
Indeed,  the relevant perturbations  turn out to correspond
to string oscillations in the  four   spatial directions transverse to 
the NS-NS five-brane.  These dimensions are  ``intrinsic'' to 
the black hole and do not depend on a choice of a superstring theory.
It is only these {\it four}   directions  that get multiplied 
by the (product of)  magnetic charges in the near-horizon region.
In terms of the free-oscillator 
description of perturbations 
 (which indeed applies  in  the near-horizon region   for large charges, 
i.e. large  level $\k \sim P $  of  the underlying   WZW model) 
this corresponds to the effective number
of degrees of freedom 
$c_{eff}= 4(1 + \ha) = 6$,   
where we have included the  contributions of superpartners 
of the four bosonic  string coordinates.
The  statistical entropy  obtained by counting 
the  supersymmetric string oscillations in the
four transverse directions  with the tension or oscillator
 level rescaled
by $\k$  exactly matches the BH entropy  \ci{me2}.

These  developments   were overshadowed  by  the advent of 
D-branes -- nonperturbative objects in string theory with Ramond-Ramond (R-R)
charges~\cite{polch95}.
Black holes with    NS-NS  charges can be mapped,  by 
duality symmetry,  onto black holes  with   R-R charges  which  
 have 
  the same  space-time metric and thus BH entropy. The  higher-dimensional 
interpretation  of these black holes  in terms of 
intersecting D-branes  lead to a counting of black hole 
quantum states that agrees precisely with the BH entropy for
 both the extreme ~\cite{strom96a} and  near-extreme 
\ci{calmal,horstr}  static, as well as rotating ~\ci{ell,all} black holes.

Recently,  Strominger~\cite{strom} (see also
~\cite{dublinbtz,SS}) has given an alternative derivation of the 
BH entropy, by again employing  the near-horizon geometry. 
The central observation is that, when embedded in a higher dimensional 
space, the near-horizon geometry contains locally  the three-dimensional 
anti de Sitter space-time ($AdS_3$), whose quantum states are described  
by a two-dimensional conformal field theory (CFT)  at the asymptotic boundary 
\ci{adsc}.
The counting of states in this CFT  is  then used to  reproduce 
the BH entropy. 
While  apparently elegant and compelling, 
this approach 
does not  provide  a 
 detailed understanding of microstates  involved 
and  recently came under some  criticism~\cite{carlip}.
The method nevertheless   reproduces the  BH entropy of the 
 static near-extreme 
black holes in
four~\ci{strom}  and
five~\ci{bl98} dimensions  as well as  that of the  corresponding
 rotating  black holes~\cite{cl98a,cl98b}.

The robustness  of these results 
 for  both  extreme and near-extreme black
holes  renewed the  interest in trying to  relate  the details of
black hole microscopics to  features of the black 
hole near-horizon geometry~\cite{strom98a,martinecads3,larsen}. 
 One of the aims  of the present  paper is to  generalize the   earlier
approach~\cite{us2,me1,me2}
to counting of  microstates for BPS-saturated
 black-holes with NS-NS charges
as perturbations of the WZW models, describing  the  curved
space-time geometry of the near-horizon region, 
 to the case 
of 
{\it non-extreme} black holes.
(An early
  attempt in that direction  
was  made in \ci{russo}.)

Our approach  is different  from 
that of~\ci{strom,bl98,cl98a,cl98b}  as it does not use   the 
 local equivalence 
of
the  Banados-Teitelboim-Zanelli 
 (BTZ) black hole~\ci{btz,hoor,kalop} and $AdS_3$,  
 and the  fact that the  asymptotic boundary  of
the latter  corresponds to   
$SL(2,R)_L\times SL(2,R)_R$ CFT~\cite{adsc}.
In our case  the microstates are counted  directly  as 
string states at the horizon
 and not at the asymptotic boundary. 
 The identification of the  underlying  
CFT  is 
more explicit  and relies on the  correspondence  (matching)
between the solitonic  and fundamental string states,
extending the discussion  in  \ci{us2,me1,me2} to non-BPS states.

The  crucial  observation    is that in the large charge limit
the  string sigma-model 
representing a  non-extremal black hole with 
NS-NS charges   is still equivalent to  a (version of)
$SL(2,R)\times SU(2)$ 
WZW  model.  Thus  the 
 near-horizon region   remains to be  effectively described 
by the 
free fundamental string whose tension is again rescaled by the 
magnetic charges.    However,  now   the black-hole 
  microstates   are identified  not only with the 
 left-moving, but also {\it  right-moving}   superconformal  string 
oscillations,
 which  can be  of the same order of magnitude. 

As a prototype example we consider the five-dimensional non-extreme rotating 
black
holes and demonstrate that   for a large NS-NS five-brane  ($P$) and 
fundmental string ($Q$)
charge this black hole is 
 described by a marginal integrable  perturbation 
of the $SL(2,R)\times SU(2)$ WZW model generated  by  the 
left- {\it and}  right-moving  Cartan 
chiral  $SU(2)$ currents 
with the perturbation parameters 
proportional to the two angular momentum components.
We shall   demonstrate  that the  non-trivial expression for 
 the BH entropy  of these 
rotating black  holes  \ci{cy96b,all} 
  can be reproduced precisely  by  identifying the  microstates as the
  left-  {\it and} right-moving oscillations  of the effective fundamental
  string.

The rest of the paper is organized as follows. In Section \ref{WZW}
 some  essential features of the  \wzw WZW model
 and its 
deformations are discussed. In Section \ref{counting} 
we consider the statistical entropy  of microstates 
represented by 
left-moving and right-moving
marginal perturbations in the four  transverse directions of the effective
string  descibed by a 
large-level WZW model. 
 Section \ref{5dbh} is devoted to  the  sigma-model corresponding
to   the non-extreme five-dimensional black holes. In 
Subsection \ref{static} the  sigma-model
for   the non-extreme  static five-dimensional 
black hole is  derived, and it is shown that 
in
the case of the large $P$ and $Q$  charges it reduces
 to the  \wzw WZW model  with   equal levels $\k=P/\a'$. In
 Subsection \ref{rotating} the \sm in 
  the case
 of the rotating black hole  with large $P$ and $Q$  is  written down 
 and is shown to be equivalent  to the  deformed
  WZW model with   the  coefficients  of the  left- and right-moving 
  deformations  proportional to the two angular momentum
  components.  We  establish the precise  relation 
between   the standard light-cone coordinates of
 the effective fundamental string and the 
canonical  coordinates   of the \wzw WZW model. 
  In Subsection \ref{entropy}  it is demonstrated that the  counting of 
 the relevant  states of the  obtained 
fundamental string   reproduces  the BH entropy of the black hole.

\section{$SL(2,R) \times SU(2)$ WZW theory and its  marginal deformations}
\label{WZW}
The central  role in our discussion will be played 
by the \wzw WZW model with equal levels $\k$ of the two factors. 
This unique model has several remarkable features. 
Since the curvatures  of the $AdS_3$ and $S^3$ factors are equal and opposite 
in sign, the leading correction to its central charge vanishes.
Moreover, in the supersymmetric case  the 
total central charge has the free-theory  value (see also  \ci{anton,us1}):
\be
c={3(\k-2)\ov (\k-2) +2} + {3(\k+2)\ov (\k+2)-2} =6.
\label{cc}
\ee
Thus multiplied, e.g., by a four-torus or $K3$, 
this CFT  represents an exact solution of the 
 $D=10$ (type II or heterotic) superstring theory describing 
\ads space supported by NS-NS two-form-tensor  and
  {\it constant} dilaton. 

This WZW theory, its orbifolds  and marginal deformations turn out to 
 describe the near-horizon regions or large charge 
limits of the basic examples
of the BPS-saturated  $D=4$  static \ci{lowe94,us1,us2} and $D=5$  static and 
rotating \ci{me1,me2} black holes
with regular horizons. 

As we shall demonstrate  in Section \ref{5dbh}, this remarkable model describes 
also 
the near-horizon limit of the {\it non-extreme} (both   static 
and  rotating)  $D=5$
black holes  with three  NS-NS charges (and two  rotation parameters).
(This approach 
can be generalized to the case of  the  four-dimensional  black holes
specified by four charges and one angular momentum parameter as  will be
discussed  elsewhere.)

 The Lagrangian of the \wzw model   can be written as\foot{The action is 
given by $I= { 1 \ov \pi \a'} \int d^2\sigma \ L$. 
We shall often set $\a'=1$ in what follows.}
\be\la{lrrg}
L= \k\left[ L_{SL(2)} + L_{SU(2)}\right]\ ,
\ee
\be
\la{slt}
L_{SL(2)} =  \four  (2 \cosh z \ \del u \bd v + 
     \del u  \bd u +   \del v  \bd v  +    \del z \bd z )\ , 
\ee
\be \la{sut} 
   L_{SU(2)}=\four (
 \del \t \bd \t  +
 \del \vp \bd \vp   +  \del \psi \bd \psi
  +  2 \cos\ \t \ \del \psi \bd \vp )
\ , \ee
where $x,u,v$ are the $AdS_3$ coordinates and 
$\t,\psi,\vp$ are the $S^3$ coordinates (Euler angles).
Note that in the  Gauss parametrisation of ${SL(2)}$ 
\be
L'_{SL(2)} =e^{-  z' } \del u'\bd v' + \four  \del z'  \bd  z'  \ . \la{gau}
\ee 
The  near-horizon limit of the  extreme $D=5$  three-charge static 
black hole
 turns out to be  represented  by the following Lagrangian \ci{me1}:
\be
 L= L''_{SL(2)} + \k  L_{SU(2)}
\ , \ \ \ \ \  \ \ \ L''_{SL(2)}
 = e^{-z} \del U \bd V +   {\td Q}Q\inv  \del U \bd U
 +  \four P \del z \bd z \ . \la{olde}
\ee 
Here $P=\k$ is the NS-NS five-brane charge, 
$Q$ is the fundamental string charge and ${\td Q}$ is  momentum along the 
string.
The radial coordinate $r$  of the black hole is expressed in 
terms of  $z$ by 
 $Q\ov r^2$= $e^z$.  The coordinates 
$(U,V)=\pm t+y$    are interpreted as the 
light-cone coordinates of the fundamental  string.

The  Lagrangians  \rf{olde},\rf{gau} and \rf{slt} are related by 
coordinate transformations.\foot{The group element of 
$SL(2,R)$ in the Gauss decomposition parametrisation is 
$
g =
 \left(\matrix{1&u\cr 0&1\cr }\right)
\left(\matrix{{e}^{{1\ov 2} z}  &  0\cr    0  & { e}^{-{1\ov 2} z}    \cr 
}\right)
\left(\matrix{ 1 &  0\cr  v  & 1 \cr }\right) , $
while 
in  the `Euler angle'
parametrisation  is 
$  g= {e}^{{i\over 2 } u \s_2 } {e}^{{1\ov 2} 
{z}\s_1} {\rm e}^{{i\over 2 }v}$.
The transformation between 
$ e^{-2x }   \del u \bd v  + n  \del u \bd u +
 \del x \bd x $ and $ 
 e^{-2x' }   \del u' \bd v'  +
 \del x' \bd x'$ is 
$ u'={1\ov 2 \sqrt n}  e^{2\sqrt n u}  ,  \
   v'= v -  \sqrt n e^{ 2x}  ,  \
   x' = x +  \sqrt n   u   .$}
These transformations
may be well-defined 
only locally if some of the coordinates are compact. This is indeed the case for
the solitonic  string  case  where 
the coordinate $y=\ha(U+V)$ along the string is compact.

Making a formal  coordinate transformation that mixes the parameters
of $SL(2,R)$ and $SU(2)$  
(but may not respect periodicities of the coordinates
and thus, in particular,   break supersymmetry \ci{old,kir})
one obtains an exact conformal model  which is a 
marginal deformation of the  original 
WZW model. This deformation
itself represents a near-horizon region of a different string solution. 
 Previous examples   were considered in \ci{us2,me1}.

A particularly  simple   deformation  which  will be  relevant in 
the present  paper
is generated by
the  Cartan $SU(2)$ chiral currents   and 
   is obtained by the following redefinition\foot{The corresponding marginal 
deformation
can be interpreted also  as a certain gauging  of 
 WZW   model.}
\be
\vp \to \vp + q_2 v  \ , \ \ \ \ \ \ \ 
\  \psi \to \psi +  q_1 u \ , 
\la{tran}\ee
where $q_1,q_2$ are two arbitrary constants.
This leads (after integrating by parts) 
to the following model
$$
L =  \four  \k  \left[ 2( \cosh z  + q_1 q_2 \cos\ \t) 
 \del u \bd v + 
    (1 + q_1^2)  \del u  \bd u +  (1 + q_2^2) 
 \del v  \bd v  +    \del z \bd z \right]
$$
\be
+ \  \ha  \k ( q_1  \del u \bar J_3  + q_2  J_3 \bd v )   +  L_{SU(2)}
\ , \la{defo}
\ee
where  
\be
J_3= \del \vp \  + \cos\ \t\ \del \psi\ , 
\ \ \ \ \ \ \ \ \
\bar J_3= \bd \psi \ +  \cos\ \t\ \bd \vp\ .
\la{carr}
\ee
If  a combination of $u$  and  $v$  is  periodic,  the transformation \rf{tran}
does not   preserve periodicities
but the model \rf{defo}  is  still  well-defined.
While locally it is 
a direct product, globally
it is  not 
for generic values of $q_1,q_2$.

In Section \ref{5dbh} we shall show 
that  essentially the same 
WZW  theory  \rf{defo}  describes the  large  $P$ and $Q$ charge limit 
of the non-extreme 
rotating $D=5$  black hole, with  $q_1,q_2$ being proportional  to the two
angular momentum parameters (see Subsection \ref{rotating}).
(Although this model 
has two   ``rotational'' parameters,  in contrast to the models
considered in \ci{me3},  it  has 
a natural   extension to finite   radial distance   which is 
asymptotically flat.)
The original   undeformed \wzw model  in the form written in 
\rf{lrrg},\rf{slt},\rf{sut}  describes the  large $P$, $Q$ charge 
limit  of the  static non-extreme 
black hole (see Subsection \ref{static}). 
We shall  then demonstrate  (see Subsection \ref{entropy}) 
that the  non-trivial expression for the 
non-extreme rotating black  hole  entropy \ci{cy96b,all} 
  can be reproduced by applying the counting of  microstates  discussed in
    Section \ref{counting}. This  generalizes the previous
  analysis in the case of
  extreme black holes
\ci{us2,me1,me2} to the non-extremal  case.

\section{Near-horizon string perturbations\\ and  statistical  entropy}
\label{counting}
In this section we shall generalize  the 
counting of  microstates of the effective fundamental string, 
 as  
initiated in \ci{us2,me1}  and further clarified   in \ci{me2}, to the   case
that along with the left-moving   marginal perturbations 
{\it includes also the right-moving marginal perturbations}.
The main idea is that in the large charge  limit  the  \wzw WZW model  describes
a fundamental string with a tension  rescaled by the NS-NS five-brane charge 
$P$, 
and whose 
relevant marginal deformations 
are  only those of 
 the four transverse 
directions to the NS-NS five-brane.  (Oscillations 
within 5-brane  are suppressed by a large charge factor 
and contribute only to subleading  terms in the entropy \ci{me2}).
 Since the charges are large, 
the conformal \sm  is weakly coupled, so one is 
 to count  string states
 in nearly-flat space.
To relate the oscillation numbers of the  fundamental
string and thus 
its entropy to the global parameters  of the  black hole 
 one  is to use the ``matching conditions'' 
 analogous to the ones in  \ci{cmp,harv,me2}.

Let us  combine the  $z$-direction of $SL(2)$ in \rf{slt},  related to the
radial coordinate $r$ of the black hole  and three  angular
 coordinates of   ($\t,\ \p , \ \psi$)
of $SU(2)$ in \rf{sut}    into  four  transverse coordinates  $x^i$.
The key point is that since we are  interested in the large charge 
or large WZW level limit,   interactions in  WZW theory are suppressed
 by $1/\k$, and thus   the count of perturbations should be essentially the 
same as in the theory
of free fields $x^i$. 
The only difference compared to the flat space case is the presence of the 
factor $\k$ in front of the kinetic term of $x^i$. 
To make this more explicit, let us 
 consider the following   sigma model  which represents the  perturbed
version of the 
model \rf{lrrg}  (cf. \ci{hor,us2}) 
\bea\non
  L&=& F(x) \del u \bd v  + K_0 \de u \bd u + M_0 \de v  \bd v
\\
&+& \  \A_i (x,u) \de u \bd x^i  +  \AA_i (x,v) \de x^i\bd v 
+   \four \k h_{ij}(x) \de x^i \bd x^j \  . 
\la{gene}
\eea 
  Here $u,v$ 
correspond to the  light-cone coordinates $(u,v)=\mp t+y$
of the 
effective  fundamental string. 
 We have included  along with the ``left-moving" perturbations, $\sim \del
 u$, also the ``right-moving" ones, $\sim \bd v$. 
 
 Since the level $\k $ is large, we may assume 
that $h_{ij} = G_{ij} + B_{ij}$ is approximately flat and  also set  $F(x) 
\approx 1$ (the large charge limit corresponds to 
 the small curvature of  $AdS_3$ space).
  As a result, the perturbations  $\A_i (x,u)$
and $\AA_i (x,v)$ are marginal to the leading order in their strength. Indeed, 
integrating out $u$ ``freezes'' $v$ and thus makes 
$\AA_i $ marginal and vice versa for $\A_i$. 
In general, there is also a constraint on their $x$-dependence,
and  the solution which is relevant in the large charge limit is 
\be\la{soo}
\A_i (x,u) \sim F(x) a_i(u)\to a_i(u)\   , \ \ \ \ \ \  \ 
\AA_i (x,v) \sim F(x) b_i(v)\to b_i(v)\ . 
\ee
Integrating out the four transverse fields $x^i$ 
in the large $\k$ limit we find
that $K_0$ and $M_0$ in \rf{gene}  are replaced by
\be
\la{rep}
K(u) = K_0 -\four  \k\inv a^2_i(u)  + O(k^{-2}) \ , 
\ \ \ \ \
M(v) = M_0 - \four \k\inv b^2_i(v)  + O(k^{-2}) \ . 
\ee
In general, there is also a  $\A\AA$-correction to 
the $\del u \bd v$ term.

Generalising the discussion in  \ci{me2}
 to include also the
right-moving oscillations,  we assume that 
the analogue of the level matching condition 
for the free fundamental string \ci{cmp,harv}
which relates  the oscillation level numbers $N_{L} $ and $N_{R}$ 
to the  background 
charges is  as follows:   the coefficients of  both  $\del u \bd u$ and
$\del v \bd v$ terms  should vanish on
 average to allow matching onto a fundamental  string source, 
i.e. 
\be
\overline{ K } =0 \ , \ \ \ \  \ \ \ \ \ \overline{ M } =0 \ , 
\ \ \ \ \ \ \ \   \overline{  f }
  \equiv {1 \ov 2\pi R} \int^{2\pi R}_0 dy \  
f(y) \ ,  \la{kyk}
\ee
where $R$ is the radius of  a compact direction along which the string is wound.

Then,  to the leading order in $1/\kappa$,  
\be
   \overline{ a^2_i}=4\k K_0  = q N_L  \ , \ \ \ \ \ \ \ \ \  
\overline{ b^2_i} = 4\k M_0=q  N_R  \ ,   \la{reee}
\ee
where  $N_L$ and $N_R$ are interpreted as the left- and the right-moving 
string oscillation numbers
and $q$ is a proportionality  constant, depending on  the  tension of the  
fundamental string
($q={\pi \ov 4 G_5  }=1$  in the units  used below). 

The statistical  entropy   should, in general, be 
 obtained by counting 
 all near-horizon  perturbations in all  possible
  directions, but 
the  oscillations  in  the  four  ``external'' spatial   dimensions  $x^i$ 
   have dominant statistical weight for large charges  ($\k \gg 1$) \ci{me2}.
Taking into account the superpartners of the four bosonic 
oscillation directions\foot{We are  assuming that the world-sheet theory is 
$(1,1)$ supersymmetric as in   
type II theory, but the same 
conclusions are true in the heterotic case \ci{me2}
(provided
the black hole solution is embedded into the heterotic 
theory in the  manifestly conformally-invariant 
``symmetric'' way).}
 so that the effective  central charge
is $c=4(1+{3\over 2})=6$,  
 the   leading term in the statistical entropy 
is then  given by 
\be
S_{stat} = 2\pi (  \sqrt {N_{L}} +  \sqrt {N_{R}}\  )\ . 
\la{ent}
\ee

\section{Sigma-model and microstates of \\
five-dimensional  non-extreme  rotating black hole}
\label{5dbh}
In this section we shall  derive the large-charge \sm 
for five-dimensional rotating  non-extreme black holes
 and count the corresponding   microstates.

\subsection{Sigma-model for  non-extreme  static black hole
}
\label{static}

The static non-extreme black hole is specified by three NS-NS charges: the
NS-NS five-brane charge $P$,  the fundamental string charge $Q$  and the 
string  momentum  ${\td Q}$.
The   five-dimensional solution  (as  found in~\cite{cy96b})
 can be   written as   a {\it six-dimensional} black
string with the  following string-frame   space-time metric:  
\be
ds_6^2= H_1\inv ( - f d{ t}^2+d{ y}^2)+   H_5 (f^{-1}dr^2 + r^2d\Omega_3)\ ,
\label{metric}
\ee
where 
\be
d\Omega_3= \textstyle{1\over 4}(d\theta^2+d\vp^2 + d\psi^2 +\ 2\cos\ \theta 
\ d\vp d\psi )\ 
\ee
is the line element of a three-sphere, 
expressed in terms of the Euler
angles. 
The  two-form field $B_{mn}$  has  the form:
\be
B_{yt}=m \sinh 2\delta_{Q}\  r^{-2}  H_1\inv -1\ , \ \ \ \ \ 
B_{\psi\varphi}={\textstyle{1\over 4} } m\sinh2\delta_P \cos\ \theta \   \,,
\ee
and the dilaton field $\Phi$ is given by:
\be 
e^{2\Phi} =H_5 H_1\inv\ . 
\ee
Here 
\be
f =1 -{{2m}\over r^2}\ , \ \ \
H_1=1 + {{2m\sinh^2\delta_{Q}}\over r^2}\ ,\ \ \
H_5 =1+{{2m\sinh^2\delta_P}\over r^2} \ . 
\ee
The momentum along the string can be introduced  by the boost transformation
along the $y$ direction, i.e.
\be
U\equiv t+y\ \to \  e^{\delta_{{\td Q}}} U \ , \ \ \ \ \ \ \ \ \ 
V\equiv -t+y\   \to \  e^{-\delta_{{\td Q}}}V\  . 
\label{boost}
\ee
Here  the  parameterization is in terms of the  three ``boost'' parameters
$\delta_{P,Q,{\td Q}}$  and the non-extremality parameter  $m$, which are 
related
to the charges and the mass of the black hole in the following way:
\bea
P &=& m\sinh 2\delta_P\ , \ \ Q= m\sinh 2\delta_{Q}\ ,\non
 \ \
{\td Q}=m\sinh2\delta_{{\td Q}}\ ,\non
\\
M &=& m(\cosh 2\delta_P+ \cosh 2\delta_{Q} + \cosh 2\delta_{{\td Q}}).
\label{mass}
\eea
We work in the Planck units where the gravitational coupling
constant in five dimensions
is $G_5={\pi\over 4}$ (for relation to conventional units see 
\ci{cl98a}). 

The string  \sm   corresponding to  this 
background 
 is of the following complicated form:
\bea \non
L &= &H_1\inv \left[\bigg(1-{m\over r^2}
(1-{\textstyle{1\over 2}}e^{-2\delta_{Q}})\bigg) \del U\bd V 
 +{m\over {2r^2}}\bigg( e^{-2\delta_{Q}}     \del V  \bd U +  
e^{2\delta_{{\td Q}}} \del U \bd U+e^{-2\delta_{{\td Q}}}
 \del V \bd V\bigg)\right]  \\
&+ &\  H_5 r^2\left[{\del r\bd r\over r^2-2m } +{\textstyle{1\over 4}}\bigg(
 \del \t \bd \t  +
 \del \vp \bd \vp   +  \del \psi \bd \psi
  +   \cos\ \t \ (\del \psi \bd \vp  +   
  \del \vp \bd \psi)    \bigg)\right]\  \non
\\ 
&+&
{\textstyle{1\over 4 } } P \cos\ \theta \
(\del \psi \bd \vp  - 
  \del \vp \bd \psi) \ , 
\la{comp}\eea
where  one should also  add
 the  dilaton term  $\sim \Phi{\cal R}^{(2)}={1\over 2}
\log ( H_1\inv H_5)  {\cal R}^{(2)}$.

In the limit of large  $P$ and $Q$ charges, i.e.  $
\delta_{P,Q}\gg 1$, 
the    
(static) non-extreme 
 $D=6$ conformal 
model  (\ref{comp})  takes the following simpler form
\bea\non
L_{static}\equiv (L)_{\delta_{P,Q\gg 1}}&=&{Q\inv } \bigg[\left(r^2- \ha(
r_+^2+r_-^2)\right) \del 
U\bd V \non \\
&+& \textstyle{1\over 4} (r_++r_-)^2e^{2\delta_{{\td Q}}} \del U \bd U +
\textstyle{1\over 4} (r_+-r_-)^2e^{-2\delta_{{\td Q}}} \del V  \bd
V \bigg] \la{lcl}  \\
&+&  P\bigg[{{r^2 \del r\bd r}\over {(r^2-r_+^2)(r^2-r_-^2)}} 
+\textstyle{1\over 4}\bigg(
 \del \t \bd \t  +
 \del \vp \bd \vp   +  \del \psi \bd \psi
  +  2 \cos\ \t \ \del \psi \bd \vp \bigg)\bigg]\ , 
 \non\eea
with the {\it constant} dilaton $e^{2\Phi}={P\over Q}$.
Here 
 \be 
 r_{+} =\sqrt {2m} \ , \ \ \ \ \ \ \ \ \ \   r_-=0 \ ,  \la{yyy}
 \ee 
  correspond to the  location of the inner
and outer horizons of the static black hole. 

The  Lagrangian (\ref{lcl})  is  indeed the  \wzw WZW model (\ref{lrrg}), 
discussed 
in Section 
\ref{WZW},
 with  the  level $\k={P}$ ! The coordinate  $z$ in (\ref{lrrg}) 
and
 the radial coordinate $r$ in (\ref{lcl}) are related  by
 \be
 {\cosh {z\over 2}}= {\sqrt{r^2- r_-^2\over {r_+^2-r_-^2}}}\  . \la{zzz}
\ee
The   coordinates  $u,v$  of the WZW model  (\ref{lrrg}) are 
proportional 
to the   light-cone coordinates $U,V$ in  \rf{lcl}
\be
u= {1\over {\sqrt{PQ}}} (r_++r_-)e^{\delta_{{\td Q}}}\ U\ ,\ 
\ \ \ \ \ \ \ \ \ \ 
v= {1\over {\sqrt{PQ}}} (r_+-r_-)e^{-\delta_{{\td Q}}}\  V\ . \la{vvv}
\ee
It is instructive to  recall  the BPS-saturated limit of 
(\ref{lcl})~\cite{me1}. This limit is obtained by taking 
 $m \to 0$  and $\delta_{P,Q,{\tilde Q}}\to \infty$ 
while keeping the charges  $P,\ Q, \ {\tilde Q}$ fixed. 
The Lagrangian (\ref{lcl})  then  becomes:
\be
L_{static}^{BPS}={Q\inv } \left(r^2 \del U\bd V + 
{\tilde Q}\del U \bd U\right)+  P\left(r^{-2} \del r\bd r
+  L_{SU(2)}\right) \ . 
 \la{lclbps}
 \ee
 After the  transformation of the radial coordinate $r\to 
z=-\log{r^2\over Q}$ 
becomes
 that of  (\ref{olde}) \cite{me1}.
 Note that the terms $\sim \del V \bd V$ 
are
 absent in the BPS limit. 

\subsection{The large charge limit of the sigma model \\ for non-extreme 
rotating black hole}
\label{rotating}
 The  $D=5$ rotating black hole  is specified, 
in addition to the three charges  and the
 mass (\ref{mass}),   by the two angular momentum parameters
\be 
{\cal J}_{L,R} = m(l_1\pm l_2) \bigg( 
\cosh\delta_P\cosh\delta_{Q}\cosh\delta_{{\td Q}} \pm
\sinh \delta_P\sinh\delta_{Q}\sinh\delta_{{\td Q}}\bigg)  \ , 
\ee
where $l_{1,2}$ are the angular momenta of the corresponding neutral black hole.

 Its Einstein-frame $D=5$  metric, dilaton  and  the antisymmetric tensor 
  field $B_{mn}$  found  in \cite{cy96b} 
are of  very   complicated form. The 
   metric of the  associated  
six-dimensional rotating black string solution
  given in \cite{cl98a} is more transparent. (By duality,  this Einstein-frame 
 metric is the same  as 
  that of the  
  six-dimensional rotating string with R-R charges.)
The  string  \sm  which  still looks rather involved, 
  simplifies dramatically  in the
  large $P$,$Q$ charge limit, i.e. $\delta_{P,Q}\gg 1$\ 
 (see eqs.(12)--(15) in \cite{cl98a} for the Einstein-frame
  metric). In this limit  it takes  
   the following form (cf. \rf{lcl}) 
\bea
L_{rot}= (L)_{\delta_{P,Q}\gg 1}
&=& 
{Q\inv } \bigg[\left(r^2- \ha(
r_+^2+r_-^2)  + \ha (r_+^2- r_-^2) q_1 q_2 \cos\ \theta \ \right) \del 
U\bd V \non\\ 
 &+&   \textstyle{1\over 4} (1 + q_1^2) 
(r_++r_-)^2 e^{2\delta_{{\td Q}}} \del U \bd U +
\textstyle{1\over 4} (1 + q_2^2) 
(r_+-r_-)^2e^{-2\delta_{{\td Q}}} \del V  \bd
V \bigg] \non\\
&-& \textstyle{1\over 2}(l_1+l_2) \sqrt{P\over Q}
e^{\delta_{\tilde Q}}\ \del U{\bar J}_3  
- \textstyle{1\over 2}(l_1-l_2)\sqrt{P\over Q}
e^{-\delta_{\tilde Q}}\ J_3\bd V  
\non\\
&+&   P\bigg[{{r^2 \del r\bd r}\over {(r^2-r_+^2)(r^2-r_-^2)}} 
+   L_{SU(2)}\bigg] \ ,  \la{def}
\eea
 with  the constant dilaton $e^{2\Phi}={P\over Q}$.  Here  
\be
r_{\pm}= {1\over 2}\bigg[ \sqrt{2m-(l_1+l_2)^2}\pm 
\sqrt{2m-(l_1-l_2)^2}\bigg]\ ,
\la{rrrp}
\ee
and  
\be
q_1= - {{l_1+l_2}\over {r_++r_-}}\ , \ \ \ \ \ \ \ \ \
  q_2=- {{l_1-l_2}\over{r_+-r_-}}\ .
\ee
The 
$SU(2)$ 
chiral 
currents $J_3, \  {\bar J}_3$ are  defined in (\ref{carr}).

Using the 
transformation  (\ref{zzz}) 
  between $r$ and $z$ 
 and   the transformation (\ref{vvv}) between
$(U,V)$ and $(u,v)$    (now with $r_{\pm}$  defined by (\ref{rrrp}))
it is easy to see that 
\rf{def}  is  precisely the 
 integrable marginal 
deformation \rf{defo} of the \wzw WZW model 
discussed in Section \ref{WZW} !

This can be seen also by noting that 
 after the  coordinate transformation~\cite{cl98a}
\be
{\vp}\to  \vp - (l_1-l_2) (PQ)^{-{1\over 2}}e^{-\delta_{\tilde Q}}
V  \ , \ \ \ 
{  \psi}\to   \psi - (l_1+l_2)(PQ)^{-{1\over 2}}e^{\delta_{\tilde Q}} U\ ,
\la{fred}\ee
i.e. the shift  \rf{tran}, 
the Lagrangian  (\ref{def}) becomes  that of (\ref{lcl}), but  now  with 
 the locations $r_\pm$ 
of the inner and outer horizons  
  related 
to the non-extremality
parameter $m$  and the angular momentum parameters $l_{1},l_{2}$ 
by \rf{rrrp}, i.e. as for  {\it rotating} black hole. 
 Thus after the formal  field 
redefinition (\ref{fred})  the sigma-model Lagrangian (\ref{def})
 can be locally written as (\ref{lcl}), i.e. as  a  direct-product
\wzw WZW model  (\ref{lrrg}) with  level $\k=P$.
Further redefinition of $r$ and $U,V$  which  
leaves the form of (\ref{lcl}) invariant but effectively replaces
$r_\pm$ in   \rf{rrrp} by the $l_1,l_2$-independent ones
in \rf{yyy} explains   why \rf{def} may be  interpreted 
also as a deformation of the static ($l_1,l_2=0$) 
WZW model (\ref{lcl}).


The BPS-saturated limit of  the  black hole solution
 is found by taking 
 $m \sim l_{1}^2\sim l_{2}^2 \to 0$,  $\delta_{P,Q,{\tilde Q}}\to \infty$,
 with  the charges  $P, Q,  {\tilde Q}$  and ${\cal J}_L$ fixed
(${\cal J}_R\to 0$). 
Then  the Lagrangian (\ref{def}) becomes \ci{me1} 
\bea \non
L_{rot}^{BPS}&=&{Q\inv } \left(r^2 \del 
U\bd V + {\tilde Q} \del U \bd U    -  {\cal J}_L \del U{\bar J}_3 
  \right) \\&+&
  P\left(r^{-2}\del r\bd r  + L_{SU(2)}
 \right) \ , 
\la{bpsr}\eea
which can be interpreted as a  marginal 
deformation of the   \sm  \rf{lclbps}  representing the static black hole
\ci{me1}.
Note that  in this case  the perturbation 
involves only  ``right''  (${\bar J}_3$) chiral $SU(2)$ Cartan current.

\subsection{Statistical  entropy}
\label{entropy}
The general formula for the BH entropy of rotating black
holes in five dimensions is~\cite{cy96b}:
\bea
S_{BH}=\ 
2\pi &m&\left[ (\prod^3_{i=1}\cosh\delta_i+\prod^3_{i=1}\sinh\delta_i)
\sqrt{2m-(l_1+l_2)^2}\right. \nonumber \\
&+&
\left. (\prod^3_{i=1}\cosh\delta_i-\prod^3_{i=1}\sinh\delta_i)
\sqrt{2m-(l_1-l_2)^2}\ \right]~.
\label{eq:macroent}
\eea
where $\delta_{1,2,3}$ is a short notation for $\delta_{P,Q,{\td Q}}$.

In the limit $\delta_{P,Q}\gg 1$ this becomes~\cite{cy96b,all}:
\bea\non
S_{BH}&=& 
\pi
\sqrt{P Q}\left[\sqrt{2m-(l_1+l_2)^2}~
e^{\delta_{{\td Q}}}
+\sqrt{2m-(l_1-l_2)^2}~e^{-\delta_{{\td Q}}}\right] \\
&=&\pi
\sqrt{P Q}\left[(r_++r_-)
e^{\delta_{{\td Q}}}
+(r_+-r_-)e^{-\delta_{{\td Q}}}\right]\ .\label{eq:ent2}
\eea
In the extreme case the entropy  takes the form \ci{ell,me1}
\be 
{ S}_{BH}^{BPS}= 
2\pi \sqrt{ P  Q  {\td Q}- {\cal J}_L^2}\ .
\la{entbps}
\ee
Our aim will be to  employ the  method of  Section \ref{counting} 
 to give a 
microscopic interpretation  of the BH entropy  (\ref{eq:ent2}). 
For  that we will need to identify
properly the  oscillation numbers 
$N_L$ and $N_R$ which   determine the statistic entropy 
(\ref{ent}).

It is instructive to recall   first   the 
 explicit expressions  one finds in 
the extreme case \ci{me1,me2}, where there is just  one
rotational parameter ${\cal J}_L$ (see (\ref{bpsr}).
In this case $M_0 \del v \bd v$ term in \rf{gene}
is absent, and so one is to ignore   the right-moving 
 ($\sim \bd v$) perturbations: their contribution 
will vanish according to the matching condition 
in \rf{rep}, i.e. 
 one counts essentially  only the 
BPS-saturated states.
For the   $D=6$ string \ 
$
{\td Q}=  {n \ov R}  , \ 
Q=  {wR } , 
$ 
where $w$ and $m$ are  integer winding and momentum numbers, $R$ is the radius
of the compact direction $y$ and as above we assume
 $G_5={\pi\over 4}, \ \a'=1$.
Special rotational perturbation  of the    $\A_i$-type  in (\ref{gene})
which is  proportional to  the chiral $SU(2)$  current
${\bar J}_3$, \ 
 $\A_i 
\sim a(u), 
 $  \ 
contains
the  constant  part, $\overline a ={\cal J}_L$ (see \ci{me2} for details).
 One finds that
the corresponding rotational parameter
should be quantised  \ci{me1}:  
$  {\cal J}_L  R P\inv {Q}\inv  = l=$integer. The quantisation of 
${Q}={w R} $
and  $P= \k$ 
implies  that the angular momentum  takes integer values 
${\cal J}_L= \k w l=j$.
Expressing  the ``left''   matching condition  in \rf{reee} (relating 
 the throat region 
perturbations to the charges)
in terms of  the integer quantum numbers one finds 
 \ci{me1,me2}
\be
N_{L}= \overline{a^2_i } = PQ{\tilde Q}-{\cal J}_L^2 = \kappa w  n -j^2 ,
\ee
so that the statistical entropy 
\rf{ent} reproduces  indeed the Bekenstein-Hawking  entropy 
(\ref{entbps}). (The  same result 
  was reached 
 using D-brane counting in \ci{ell}.)

Let us  now  turn to the discussion of the entropy in the non-extreme
 rotating case
 (with large $P$ and $Q$ charges). Here there are {\it two} angular momenta,
and  the 
Lagrangian (\ref{def})  contains 
deformation terms of both chiralities. 
 Similar 
 left-moving ($\sim
\del u$) {\it and}  right-moving  ($\sim \bd v$)
terms will  appear in the  
perturbed version \rf{gene}
of  (\ref{def}). 
The conditions of matching onto the string source 
\rf{kyk},\rf{reee}
will  then 
 determine $N_L$ and $N_R$.

As in the extreme case,  it is useful to 
 interpret  (\ref{def})  as a  special case of the  general  
deformation \rf{gene} of the static model   \rf{lcl} by assuming that the 
 special rotational 
perturbations  among 
 $\A_i$  and ${\bar \A}_i$ terms  in (\ref{gene}),  which  are proportional 
to  the chiral $SU(2)$ currents ${\bar J}_3$ and
$J_3$, respectively, 
 ${\A_i} 
\sim a (u)
$,    ${\td  \A}_i 
\sim b (v), 
$  
contain  the constant parts
 $\overline a= {\cal J}_L$ and $\overline b = {\cal J}_R$.

From   the ``left''   and ``right'' matching  conditions
  in \rf{reee} one finds\foot{Note that after integrating out 
  the deformations  in (\ref{def}) as discussed in Section III,
 the Lagrangian becomes
  that of  (\ref{lcl})  with  coefficients in front of 
 $\del U \bd U$   and $ \del V \bd V$    being  proportional to 
the coefficients in the l.h.s. of \rf{reee}, i.e. to 
$N_L$ and
 $N_R$, respectively. } 
\be
N_{L} =\four P Q (r_++r_-)^2
e^{2\delta_{{\td Q}}}\ , \ \ \ \ \ \ \ \ \ \ 
N_R=\four PQ (r_+-r_-)^2e^{-2\delta_{{\td Q}}}\ . 
\ee
Inserting these expressions  into the statistical entropy (\ref{ent}),  
we 
reproduce  precisely the (large charge limit of)
Bekenstein-Hawking  entropy \rf{eq:ent2} of the 
non-extreme rotating black hole.

 \section*{Acknowledgments}

 M.C. would like to thank F. Larsen,  R. Myers,  A. Peet,  J. Polchinski, S.
 Wadia and 
other
 participants of the  Duality program at ITP, Santa Barbara, 
 for discussions.
The work was supported in part by DOE grant DOE-FG02-95ER40893 (M.C.), 
and  in part by 
 PPARC and 
  the European Commission  TMR programme  grant ERBFMRX-CT96-0045 (A.A.T.).  
The work at the ITP was further supported by the NSF under grant 
PHY94-07194.



\end{document}